\documentclass[namedreferences]{solarphysics}

\pdfoutput=1
\usepackage[hyperref,optionalrh,showbiblabels]{spr-sola-addons} 
\usepackage{epsfig,epstopdf}
\usepackage{graphicx}        
\usepackage{color}           
\usepackage{breakurl}        
\usepackage{tabularx}
\usepackage{appendix}

\newcommand {\kms} {km\,s$^{-1}$}
\newcommand{\arcsec} {$^{\prime\prime}$}

\graphicspath{{figs/}}
\chardef\us=`\_

\chardef\us=`\_

\begin{document}

\begin{article}
\begin{opening}

\title{On the relation between transition region network jets and coronal plumes}

\author[email={qiyq@mail.sdu.edu.cn}]{\inits{Y.Q.}\fnm{Youqian}~\lnm{Qi}}
\author{\fnm{Zhenghua}~\lnm{Huang}}
\author[corref,email={xld@sdu.edu.cn}]{\inits{L.D.}\fnm{Lidong}~\lnm{Xia}}
\author{\fnm{Bo}~\lnm{Li}}
\author{\fnm{Hui}~\lnm{Fu}}
\author{\fnm{Weixin}~\lnm{Liu}}
\author{\fnm{Mingzhe}~\lnm{Sun}}
\author{\fnm{Zhenyong}~\lnm{Hou}}
\address{Shandong Provincial Key Laboratory of Optical Astronomy and Solar-Terrestrial Environment, Institute of Space Sciences, Shandong University, Weihai, 264209 Shandong, China}

\runningauthor{Y.Q.\ Qi \emph{et al.}}
\runningtitle{Relation between TR network jets and coronal plumes}

\begin{abstract}
Both coronal plumes and network jets are rooted in network lanes.
The relationship between the two, however, has yet to be addressed.
For this purpose, we perform an observational analysis using
    images acquired with the Atmospheric Imaging Assembly (AIA) 171\,\AA\ passband to follow the evolution of coronal plumes,
    the observations taken by the Interface Region Imaging Spectrograph (IRIS) slit-jaw 1330\,\AA\ to study the network jets,
    and the line-of-sight magnetograms taken by the Helioseismic and Magnetic Imager (HMI) to overview the the photospheric magnetic features in the regions.
Four regions in the network lanes are identified, and labeled ``R1--R4''.
We find that coronal plumes are clearly seen only in ``R1''\&''R2'' but not in ``R3''\&``R4'',
    even though network jets abound in all these regions.
Furthermore, while magnetic features in all these regions are dominated by positive polarity, they are more compact (suggesting stronger convergence) in ``R1''\&``R2'' than that in ``R3''\&``R4''.
We develop an automated method to identify and track the network jets in the regions.
We find that the network jets rooted in ``R1''\&``R2'' are higher and faster than that in ``R3''\&``R4'',
    indicating that network regions producing stronger coronal plumes also tend to produce more dynamic network jets.
We suggest that the stronger convergence in ``R1''\&``R2'' might provide a condition for faster shocks and/or more small-scale magnetic reconnection events that power more dynamic network jets and coronal plumes.
\end{abstract}
\end{opening}

\section{Introduction} \label{sec:intro}
\par
Coronal plumes (or rays) are bright ray-like features in the corona that could extend to tens of solar radii\,\citep[e.g.][]{1997SoPh..175..393D,2001ApJ...546..569D}.
They were first found in polar coronal holes during the eclipse more than 100 years ago\,\citep[see descriptions of the observations in][]{1950BAN....11..150V}.
Coronal plumes have also identified in UV, EUV and X-ray while such observations are available in the space era.
With {\it Skylab} observations at Mg\,{\sc ix}, \citet{1995ApJ...452..457W} first identified plume-like features in the low-latitude coronal holes,
and they proposed that coronal plumes are not unique in the polar region but may occur in open field region at any latitudes on the Sun.
This has been confirmed by observations from SOHO, TRACE, Hinode, STEREO and SDO\,\citep[e.g.,][]{1999JGR...104.9753D,2003A&A...398..743D,2008SoPh..249...17W,2009ApJ...700..292F,2011ApJ...736..130T,2011SCPMA..54.1906Y,2018ApJ...862...18D}.
The electron density of coronal plumes is 3--8 times larger than that in the inter-plume region\,\citep[][and references therein]{2011A&ARv..19...35W},
and their lifetimes range from $\sim$20\,hours\,\citep{1997ESASP.404..487L} to days\,\citep{1991sia..book.1087W,1999A&A...350..286Y,2001ApJ...560..490D}.
Since coronal plumes are always present in coronal holes where are the major source of fast solar wind\,\citep[e.g.][]{2003A&A...399L...5X,2004A&A...424.1025X,2009LRSP....6....3C,2012A&A...545A..67M,2015SoPh..290.1399F},
the connection between coronal plumes and the solar wind has been studied intensively.
Some have indicated that they might feed sufficient plasma and energy into the fast solar wind\,\citep[e.g.][]{1994SSRv...70..391V,2003ApJ...589..623G,2005ApJ...635L.185G,2010ApJ...709L..88T,2011ApJ...736..130T,2014ApJ...794..109F,2015ApJ...806..273L},
although they might be not the major source\,\citep[e.g.][]{1994ApJ...435L.153W,1995GeoRL..22.1465H,1999Sci...283..810H,2000A&A...359L...1P,2000ApJ...531L..79G,2000A&A...353..749W,2003ApJ...588..566T}.
More details about the history of the observations of coronal plumes could be found in recent reviews\,\citep{2011A&ARv..19...35W,2015LRSP...12....7P}.

\par
Coronal plumes are classified into two types, beam plumes and network plumes\,\citep{2009ApJ...700..551G}.
Beam plumes have (quasi-)cylindrical shape with a base diameter of $\sim$30\,Mm\,\citep[e.g.][]{1997SoPh..175..393D},
and their footpoints normally correspond to coronal bright points\,\citep[e.g.][]{1998ApJ...501L.145W,2008SoPh..249...17W,2014ApJ...793...86P}.
Network plumes (also named ``curtain plumes'') are made up of many faint structures that are aligned in a curtain shape\,\citep{2003ApJ...589..623G}.
Network plumes are rooted in regions along the edge of supergranular boundaries that are corresponding to network lanes in the chromosphere\,\citep{2009ApJ...700..551G,Rincon2018}.

\par
No matter which type coronal plumes belong to they are associated with bright network lanes,  where the associated magnetic features are dominated by unipolarity\,\citep[e.g.][]{1968SoPh....3..321N,1995ApJ...452..457W,1997ApJ...484L..75W,1997SoPh..175..393D,2008SoPh..249...17W,2016ApJ...818..203W,2018ApJ...861..111A}.
Recently, by investigating the evolution of the magnetic features associated with tens of coronal plumes as observed by HMI, \citet{2016ApJ...818..203W} found that coronal plumes form where unipolar network elements inside coronal holes converge to form dense clumps and fade as the clumps disperse again.
Similar results have been confirmed by \citet{2018ApJ...861..111A}, who also make a quantitative analysis and found that coronal plume appear when convergence of the associated magnetic flux surpasses a flux density of $\sim$200--600\,Mx\,cm$^{-2}$.
Along with the dominant polarity, small bipolar features also frequently appear nearby and might be cancelling with the dominant unipolarity\,\citep[e.g.,][]{1995ApJ...452..457W,1997ApJ...484L..75W,2008SoPh..249...17W}.
Based on the behaviours of magnetic features in the footpoints of the coronal plumes,
\citet{1995ApJ...452..457W} proposed a magnetic reconnection scenario,
in which reconnection occurs between open fields (corresponding to the dominant unipolarity) and nearby small-scale loops (corresponding to the small bipolar features).
Energy dissipation due to such magnetic reconnection process should take place in the base of corona\,\citep{1994ApJ...435L.153W}.
The released energy in the magnetic reconnection might conduct downward to the chromosphere and results in plasma evaporation that maintains the high density in the coronal plumes\,\citep{1998ApJ...501L.145W}.

\par
Many studies have focused on the connection between coronal plumes and other dynamic phenomena in the solar atmosphere.
For coronal plumes with coronal bright points persisting at their bases, studies found that they appear several hours after the coronal bright points first appeared and fade away hours after the coronal bright points disappeared\,\citep{1998ApJ...501L.145W,2008SoPh..249...17W,2014ApJ...793...86P}.
While coronal jets are also closely linked to bipolar regions with one dominant polarity\,\citep[e.g.][]{2010ApJ...710.1806D,2012A&A...548A..62H,2015Natur.523..437S,2016ApJ...832L...7P,navdeep2018,2018ApJ...868L..27P}, the connection between coronal jets and coronal plumes has also been investigated.
Using white light observations during the eclipse data and EIT 195\,\AA\ data from SOHO, \citet{1999SoPh..190..185L} reported that the polar coronal plume was disturbed by the jet with a speed of $\sim$200\,km\,s$^{-1}$ embedded therein.
\citet{2008ApJ...682L.137R} showed that over 90\% of the 28 jets are associated with plumes,
and they also found that 70\% of those plume-related jets are followed by plume haze occurring minutes to hours later while the rest of the jets result in brightness enhancement in the pre-existed plumes.
In more recent studies, \citet{2014ApJ...787..118R} and \citet{2018ApJ...868L..27P} confirmed the close relationship between coronal plumes and coronal jets,
and they further found that a large number of small jets and transient bright points frequently occurring in the bases of coronal plumes could be the main energy source for coronal plumes.

\par
Recently, using the high resolution observations of the transition region from the Interface Region Imaging Spectrograph (IRIS),
\citet{2014Sci...346A.315T} discovered that networks in the upper chromosphere and transition region of coronal holes are occupied by numerous small-scale jets (hereafter, network jets).
Network jets having lifetimes of 20--80\,s and speeds of 80--250\,km\,s$^{-1}$ are prevalent in the network regions.
They have been suggested to be powered by magnetic reconnection and many of them were found to be heated to at least 10$^5$\,K.
Network jets have also been found in quiet-sun region, which are slower and shorter than those in the coronal holes\,\citep{2016SoPh..291.1129N,2018A&A...616A..99K}.

\par
Since both coronal plumes and network jets are rooted in the networks and both of them might be powered by magnetic reconnection,
One would naturally ask a question what the relationship between the two phenomena are.
Targeting on this question, here we show a set of IRIS and SDO coordinated observations of a few network regions,
where present different dynamics of coronal plumes and network jets.
We develope an automatic method to identify and track jets in the network regions.
With this method, we obtain birth rates, lifetimes, lengths and speeds of the network jets.
We then compare these parameters depending on different part of the network regions.

\par
The paper is organised as follows.
The observations and methodology are described in Section~\ref{sec:obser}.
The results are shown in Section~\ref{sec:results} and Section\,\ref{sec:resstat}.
The discussion  is given in Section\,\ref{sect:discussion} and  the conclusions are given in Section~\ref{sect:conclusions}.

\section{Observations and Data Analysis}
\label{sec:obser}
\subsection{Details of Observations\label{subsec:data}}

\par
The data were taken on 2015 December 4 from 01:21\,UT to 02:19\,UT.
They were collected by IRIS~\citep{2014SoPh..289.2733D} and
the Atmospheric Imaging Assembly\,\citep[AIA,][]{2012SoPh..275...17L} and the Helioseismic and Magnetic Imager\,\citep[HMI,][]{2012SoPh..275..229S} aboard the Solar Dynamics Observatory\,\citep[SDO,][]{2012SoPh..275....3P}.

\par
IRIS was operated in a very large sit-and-stare mode and the IRIS slit-jaw (SJ) imager was observing only at 1330\,\AA\ passband with a cadence of 9\,s.
The spatial scale of the IRIS SJ images is 0.167\arcsec$\times$0.167\arcsec.
In order to reduce the telemetry load, the data had been binned by $2$\,pixels$\times$2\,pixels
and thus the spatial scale of each grid of the data array is 0.334\arcsec$\times$0.334\arcsec.
The level 2 IRIS data are used and no further calibration is required.

\par
The AIA and HMI data were downloaded from JSOC.
The AIA data include that taken in 1600\,\AA\ and 171\,\AA\ passbands.
The cadences of the AIA 1600\,\AA\ and 171\,\AA\ data are 24\,s and 12\,s, respectively.
The pixel size of the AIA data is 0.6\arcsec$\times$0.6\arcsec.
The HMI line-of-sight magnetograms with a cadence of 45\,s and a pixel size of 0.6\arcsec$\times$0.6\arcsec are used.
The AIA and HMI data are prepared with standard procedures provided by the instrument teams,
and the level 1.5 data are analysed.

\par
The images taken from different passbands are aligned using several referent features in the images.
We first align IRIS 1330\,\AA\ images to AIA 1600\,\AA, and then HMI magnetograms to AIA 1600\,\AA.
Although images from different passbands of AIA have been aligned each other by the data processing pipeline, we also check the alignment between 1600\,\AA\ and 171\,\AA\ using referent features and found an offset of 1\arcsec\ at both Solar\_X and Solar\_Y directions.

\par
\begin{figure*}[!ht]
\centering
\includegraphics[trim=0cm 3cm 0cm 4cm,clip=,width=\textwidth]{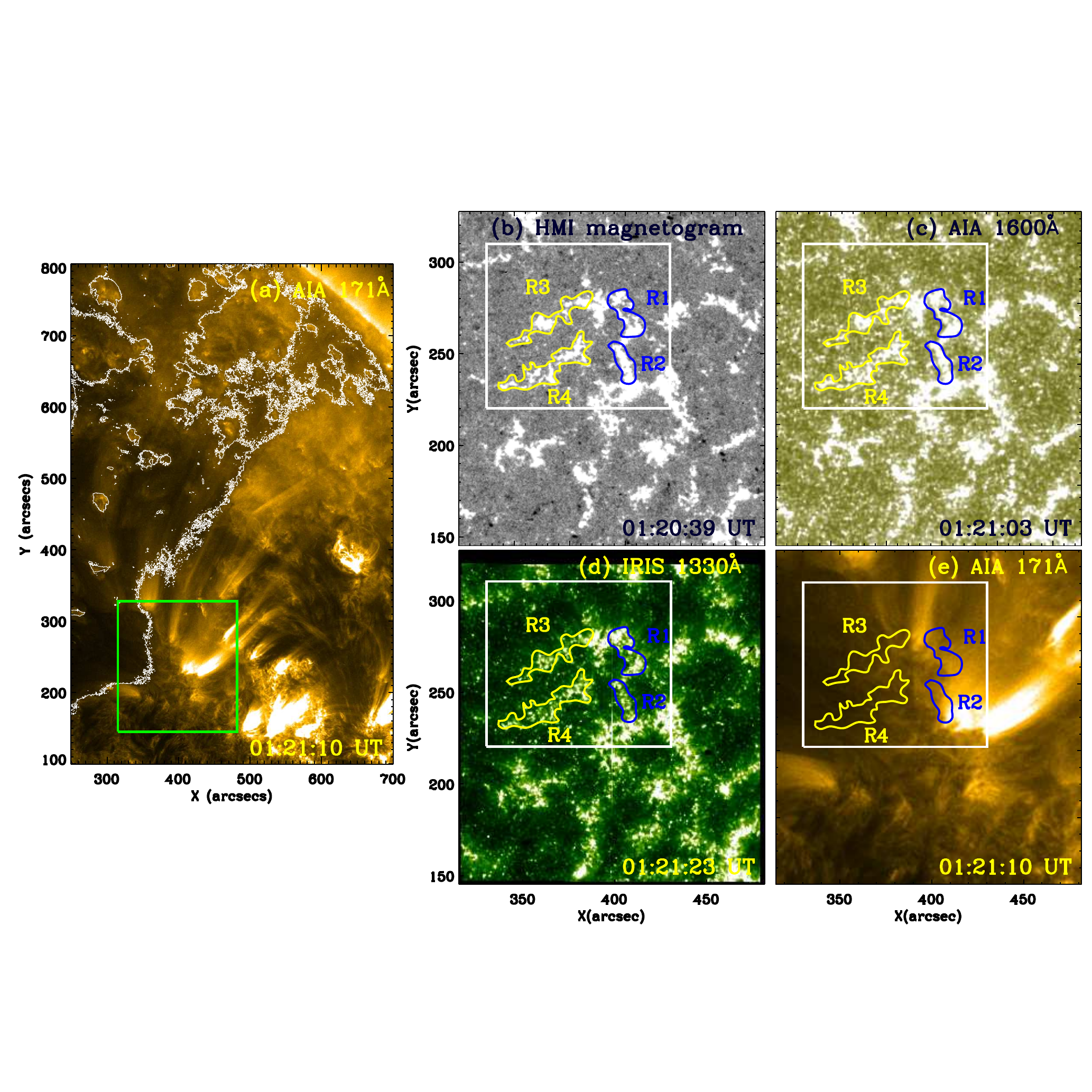}
\caption{The context images for regions studied in the present. (a) The AIA 171\,\AA\ image giving an overview of the coronal structures in and around the studied regions.
The contours (white lines) outline the boundaries of the coronal holes determined in the AIA 193\,\AA\ image.
The region enclosed by the rectangle (green lines) is zoomed-in in panels (b--e),
where show HMI magnetic features (b), AIA 1600\,\AA\ image (c), IRIS SJ 1330\,\AA\ image (d) and AIA 171\,\AA\ image (e).
The rectangles (white lines) in panels (b--e) give the field-of-view used to jet identification and tracking.
The regions outlined by blue lines with the marks of ``R1'' and ``R2'' are the network regions where coronal plumes are clearly shown.
The regions outlined by yellow lines with the marks of ``R3'' and ``R4'' are the network regions where coronal plumes are hardly seen.
}
\label{figfov}
\end{figure*}

In Figure\,\ref{figfov}, we show the context of the region-of-interest taken with AIA, HMI and IRIS around 01:21\,UT.
The region is at the boundaries of an on-disk coronal hole (see Figure\,\ref{figfov}a).
We can see that the region consists of a cluster of positive magnetic features (Figure\,\ref{figfov}b), which are aligned in a typical magnetic structure of network regions\,\citep[e.g.][]{2003A&A...399L...5X,2004A&A...424.1025X}.
The dominant polarities are in line with the coronal hole seen in the AIA data.
The network regions could be clearly seen in AIA 1600\,\AA\ and IRIS SJ 1330\,\AA\ images (see the bright lanes in Figures\,\ref{figfov}c\&d).
Many jets rooted in the network lanes are found, and they can be clearly seen in the IRIS SJ 1330\,\AA\ images.
In AIA 171\,\AA\ images (Figures\,\ref{figfov}a\&e), we can see that a set of plume-like features are rooted in some of the region.

\subsection{Network jet identification and tracking}
\label{subsec:aaptinj}

\par
Because network jets are extremely abundant in network regions, they demand an automatic method to be identified and tracked.
Here we develop an automatic algorithm for network jet identification and tracking.
In this method, if a feature is brighter than 2.5 times of background and extended for more than 4\arcsec~from the base near network lane, it is considered as a jet-like feature.
An identified jet-like feature is then traced back and forth in time to obtain its evolution.
The height of a jet-like feature is determined by the base and the point at its extending direction where its brightness drops below 2.5 times of the background.
If a jet-like feature can be recognised in more than one frames and its heights are changing with time, it is considered as a network jet.
By determining the heights of a jet changing with time, its speed could be obtained.
More details of the algorithm could be found in the Appendix.

\section{Dynamics in the network regions} \label{sec:results}

\par
In Figure\,\ref{figfov} and the associated animation, we show the evolution of the region seen in IRIS SJ 1330\,\AA\ and AIA 171\,\AA.
In the IRIS SJ 1330\,\AA\ images, we can see that jets are abundant in the network lanes.
The network jets are fine and dynamic, which are the typical characteristics of these phenomena\,\citep{2014Sci...346A.315T}.
In regions ``R1''--``R4'', the network jets ejected in directions almost parallel to each other, which allows our automatic algorithm to be used (see the Appendix for details).
In AIA 171\,\AA\ images, it is clear that plume-like features are rooted in the regions of ``R1'' and ``R2'', while they are hardly seen (or too weak to be seen) in the regions of ``R3'' and ``R4''.
It shows that the network regions of ``R1'' and ``R2'' are much brighter than ``R3'' and ``R4''.
In particular, the average irradiance of ``R1''\&``R2'' seen in IRIS SJ 1330\,\AA\ is about 2--4 times of that in ``R3''\&``R4''.

\par
As shown in Figure\,\ref{figfov} and the animations, network jets rooted in the studied regions are very dynamic.
In AIA 171\,\AA\ images, we can also observed many disturbances in the coronal plumes.
We speculate that a number of network jets were resulting in disturbances in the coronal plumes.
Because of the complex background in both transition region and coronal images, in most of the cases we cannot identify one-to-one relation between transition region activities and the coronal plume disturbances.

\par
The HMI magnetograms show that the regions are dominated by positive polarities (Figure\,\ref{figfov}b).
In agreement with \citet{2016ApJ...818..203W} and \citet{2018ApJ...861..111A}, the regions with clear coronal plumes (``R1''\&``R2'') have magnetic features more compact than that in ``R3''\&``R4''.
In most of the regions during the observing period, very few negative polarities were seen in the regions.
This indicates that magnetic features with opposite polarity to the major one, if existed, should have sizes/lifetimes under the resolution or strengths under the resolving power\,\citep[i.e. $\sim$10\,Mx\,cm$^{-2}$ in the present case,][]{2012SoPh..279..295L}.
To investigate this issue, one will need higher resolution data, which might be provided by the Goode Solar Telescope\,\citep[GST,][]{2010AN....331..620G,2010AN....331..636C}, the forthcoming Daniel K. Inouye Solar Telescope (DKIST) and the planning Chinese Advanced Solar Observatories -- Ground-based (ASO-G).
Occasionally, we can also observe that small negative polarities appear nearby the dominant positive ones and immediately disappear within around 1 minute (i.e. close to the HMI temporal resolution), thus
we cannot link such magnetic activities to any particular network jets.

\par
\begin{figure*}[!ht]
\centering
\includegraphics[trim=0cm 3cm 0cm 10cm,clip,width=\textwidth]{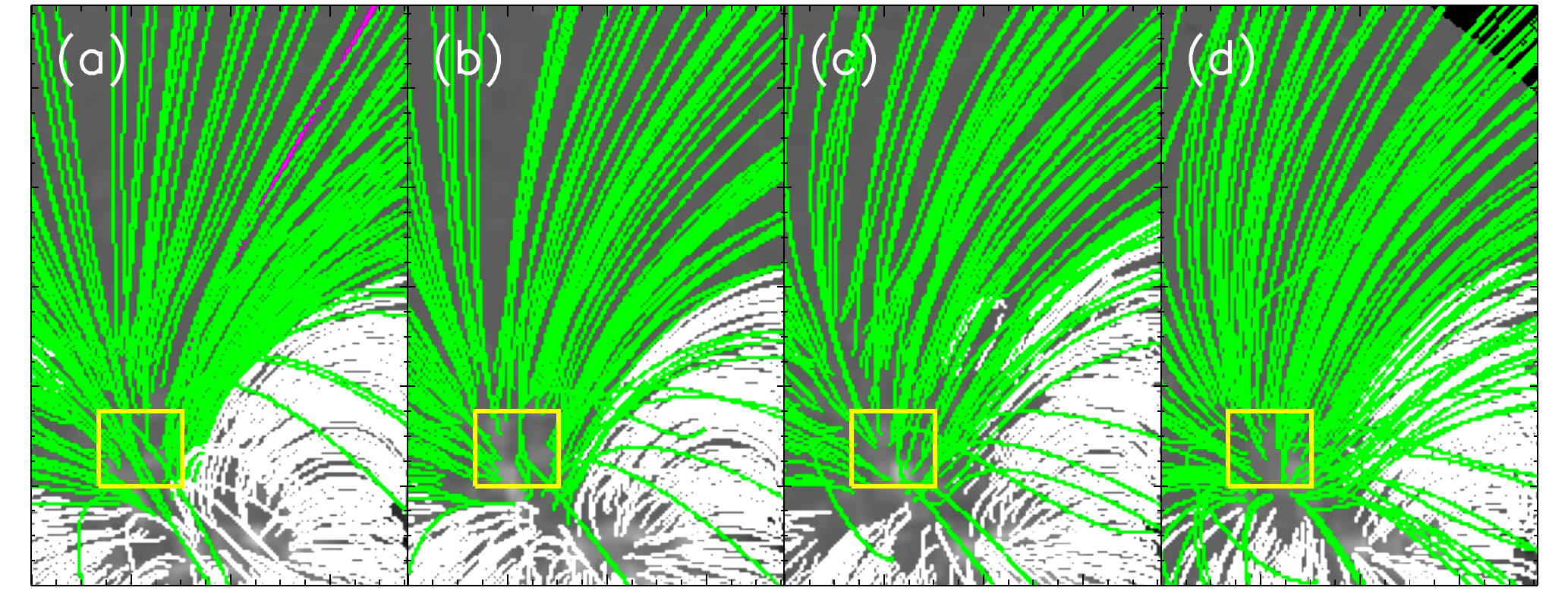}
\caption{The magnetic field lines in the regions derived from the potential field extrapolations based on data taken on December 1 (a), December 2 (b), December 3 (c) and December 4 (d).
The green and purple lines are representative of open field lines and the white ones are close field lines.
The yellow rectangles indicate the studied region as shown in Figure\,\ref{figfov}b--e.}
\label{figfield}
\end{figure*}

We examine the magnetic topology of the region using full-disk potential field extrapolation provided by the PFSS package\,\citep{2003SoPh..212..165S} in the {\it solarsoft}.
In Figure\,\ref{figfield}, we display the potential field extrapolation of the region based on observations taken on the days from December 1 to December 4, which are tracking the region from the east hemisphere to the west hemisphere.
It is clear that the region of interest is dominated by open field lines,
and the open fields could be seen at least three days before our observations.
The open topology of magnetic field is in agreement with coronal hole observed in the AIA coronal passbands,
and it provides a condition for birth of coronal plumes and network jets.
As described above, however, some network lanes in this open field region are rich in coronal plumes but some others not while both are rich in network jets.
In the following, we will compare the dynamics of the network jets in ``R1''--``R4'', and investigate the possible relation between coronal plumes and network jets.

\section{Statistical analysis of the network jets}\label{sec:resstat}

\par
In ``R1''--``R4'', we identify 1293 network jets, of which 619 located in regions ``R1'' and ``R2'' and the rest 674 located in regions ``R3'' and ``R4''.
Based on their coronal plume activities, in the statistics we group network jets from ``R1'' and ``R2'' as one category, and ``R3'' and ``R4'' as another category.
The birthrates of network jets in these two types of regions are $7.8\times10^{-16}$\,m$^{-2}$\,s$^{-1}$ (``R1''\& ``R2'') and $4.7\times10^{-16}$\,m$^{-2}$\,s$^{-1}$ (``R3''\& ``R4''), respectively.
Please note that these birthrates are the bottom limits because we only take into account the network jets that allow speed calculations (see the Appendix for detail).
In Figure\,\ref{figstatres},
we give the statistical analysis of three parameters (lifetime, height and speed) of the network jets in the two categories of regions.

\par
In the regions rich in coronal plumes (i.e. ``R1''\&"R2''), the average lifetimes, heights and speeds of all 619 network jets are 45.6\,s with a standard deviation ($1\sigma$) of 35\,s, 8.1\arcsec with $1\sigma$ of 1.6\arcsec\ and 131\,\kms\ with $1\sigma$ of 64\,\kms, respectively.
In the regions poor in coronal plumes (i.e. ``R3''\&"R4''), the average lifetimes, heights and speeds of all 674 network jets are 50.2\,s with $1\sigma$ of 35.4\,s, 5.5\arcsec with $1\sigma$ of 1.9\arcsec\ and 89\,\kms\ with $1\sigma$ of 45\,\kms, respectively.
In average, the network jets from regions rich in coronal plumes are higher and faster than that from the regions poor in coronal plumes.
However, we also see that the distributions of each parameter from the two kinds of regions are largely overlapped.
The obtained values of heights and speeds are spreading in the ranges consistent with the measurements in the previous studies\,\citep{2014Sci...346A.315T,2016SoPh..291.1129N,2018A&A...616A..99K}, but the mean and most probable values measured here are smaller.
This could be resulted from the line-of-sight effect since the region studied here is closer to the disk center.
Although the obtained lifetimes are generally in agreement with the previous studies\,\citep{2014Sci...346A.315T,2016SoPh..291.1129N,2018A&A...616A..99K}, there is a large portion showing lifetime of 18\,s (i.e. two frames in the time series).
This kind of short-lifetime jets were not included in the studies of  \citet{2014Sci...346A.315T} and \citet{2016SoPh..291.1129N} due to the temporal resolutions of their data, but they appear to be common\,\citep[see e.g.][]{2017ApJ...849L...7D,2017Sci...356.1269M}.

\par
In the speed regime, the histogram show two peaks at $\sim50$\,\kms\ and $\sim110$\,\kms\ with a clear division at $\sim90$\,\kms.
Although with less samples, such speed distribution has also been seen in that given in \citet{2014Sci...346A.315T}, but with different peak speeds at $\sim110$\,\kms\ and $\sim160$\,\kms\ and a division at $\sim140$\,\kms.
Such difference between their results and ours can be understood if we take into account the line-of-sight effect.
The questions are whether such two-peak distribution is universal and whether this two-peak distribution indicates different species of network jets.
These could be an interesting topic for future studies using higher cadence data and a larger number of samples.

\par
The distribution histograms of the lifetimes, heights and speeds of the network jets are overlapped with those of spicules that also originated from network regions\,\citep[see e.g.][]{2000SoPh..196...79S,2005A&A...438.1115X,2007PASJ...59S.655D,2012ApJ...750...16Z,0004-637X-759-1-18,2014ApJ...792L..15P}.
We speculate that a part of the network jets identified here are responses of spicules in the IRIS SJ 1330\,\AA\ passband.
To compare with the speed histogram of spicules\,\citep{2007PASJ...59S.655D,0004-637X-759-1-18}, that of the network jets much bias toward more than 100\,\kms.

\section{Discussions}
\label{sect:discussion}
Since both network jets and coronal network plumes originate in network regions,
the connection between them should be answered.
In the present study, we made a statistical analysis of a few parameters of network jets in a few network regions (``R1''--``R4''), in which coronal  plumes are clearly seen in ``R1'' and ``R2'' but almost invisible in ``R3'' and ``R4''.
We found that the network jets in ``R1'' and ``R2'' are statistically higher and faster than that in ``R3'' and ``R4''.
These observational results indicate more dynamic and energetic nature of network jets in the regions where coronal plumes are clearly seen.
It suggests that the network regions of ``R1'' and ``R2'' could provide a condition in favor of both more energetic network jets and stronger coronal plumes.

\par
While the unipolar features in ``R1'' and ``R2'' are more compact (implying a higher degree of convergence), it provides stronger compression in the regions.
The compression in the footpoints of the magnetic flux tubes might produce shocks that could feed mass and energy to the higher solar atmosphere.
The shocks have to be strong and dense enough to power plasma to produce coronal plumes in the region.
In this scenario, one would expect that the network jets are also powered by shocks and therefore many network jets should directly feed mass and energy to coronal plumes.
This will require a further study using data targeting on regions with cleaner background (e.g. polar regions), which allows identify one-to-one relation between network jets and coronal plumes.
The higher degree of compression might also produce more complicate shearing motions in the footpoints of magnetic flux tubes rooted in the region.
The complex shearing motions could generate a complexity of magnetic topology above the photosphere.
Such complexity of magnetic topology includes magnetic braids\,\citep{1983ApJ...264..635P,1983ApJ...264..642P}.
As a result of magnetic braiding, magnetic reconnections occur and could heat the plasma to coronal temperature and accelerate the plasma to more than a hundred kilometer per second\,\citep{2013Natur.493..501C,2018ApJ...854...80H}.

\par
If there were mixed polarities in the region, the higher degree of convergence might directly drive more magnetic reconnection to occur and thus feed more mass and energy to higher solar atmosphere\,\citep{2010A&A...510A..40H}.
In the present cases, although there are opposite polarities found nearby the dominant polarity, they are too rare to agree with the occurrence of the network jets.
However, it is possible that many opposite polarities cannot be resolved with the current data, and higher resolution data provided by GST and DKIST might shed light on this problem.

\par
In coronal plumes, propagating disturbances are usually seen\,\citep[see e.g.][etc.]{1998ApJ...501L.217D,2003A&A...404L..37M,2007Sci...318.1574D,2009A&A...503L..25W,2010ApJ...722.1013D,2011ApJ...738...18T,2011ApJ...736..130T,2012ApJ...759..144T}.
It has been reported that spicules\,\citep{2011Sci...331...55D,2015ApJ...807...71P,2015ApJ...809L..17J,2015ApJ...815L..16S,2016RAA....16...93J} and/or shocks\,\citep{2018ApJ...855...65H} could be the possible sources of the propagating disturbances in the coronal plumes.
In the present study, the observations show many propagating disturbances along the coronal plumes.
We can only speculate from the animation that some of the network jets should have directly linked to the propagating disturbances in the coronal plumes.
However, because of the complex background emission, we were not able to identify one-to-one corresponding between such disturbances and network jets.
Whether network jets can directly trigger propagating disturbances in the coronal plumes remains open.

\section{Conclusions}\label{sect:conclusions}
In the present study, we studied activities in four regions of network lanes, in which coronal plumes (viewed in AIA 171\,\AA\ passband) are clearly seen rooting in two (``R1''\& ``R2'') of the regions but are very fade in the other two (``R3''\& ``R4'').
In all these regions, network jets seen in IRIS SJ 1330\,\AA\ passband are abundant.
Positive polarities are dominant in these regions and negative polarities could only be seen occasionally.
The positive polarities are more compact in ``R1''\& ``R2'', suggesting higher degree of convergence.
The average irradiance of ``R1''\& ``R2'' seen in IRIS SJ 1330\,\AA\ is 2--4 times of that in ``R3''\& ``R4''.

\par
We developed an automatic method to identify and trace network jets in these regions.
With the method, we identified and traced 619 network jets in ``R1''\& ``R2'' and 674 in ``R3''\& ``R4''.
With these samples, we carried out statistical analyses of lifetimes, heights and speeds of the network jets in ``R1''\& ``R2'' and ``R3''\& ``R4'', respectively.
The lifetimes, heights and speeds of the network jets in ``R1''\& ``R2'' are 45.6\,s with a standard deviation ($1\sigma$) of 35\,s, 8.1\arcsec with $1\sigma$ of 1.6\arcsec\ and 131\,\kms\ with $1\sigma$ of 64\,\kms, respectively.
In ``R3''\&``R4'', the average lifetimes, heights and speeds of the network jets are 50.2\,s with $1\sigma$ of 35.4\,s, 5.5\arcsec with $1\sigma$ of 1.9\arcsec\ and 89\,\kms\ with $1\sigma$ of 45\,\kms, respectively.
These results show that the network jets are in-average higher and faster (i.e. more dynamic) in the regions with visible coronal plumes than that without clear coronal plumes.
We suggest that the convergence motions in the base of the network regions could build up energy and the energy could be released to the higher solar atmosphere though shocks and/or small-scale (under the current resolution) magnetic reconnections.

\acknowledgments
{\it Acknowledgments:}
The research is supported by National Natural Science Foundation of China (U1831112, 41627806,41604147, 41474150, 41404135).
Z.H. thanks the Young Scholar Program of Shandong University, Weihai (2017WHWLJH07).
We acknowledge Dr. Hui Tian, Dr. Tanmoy Samanta and Prof. Rob Rutten for fruitful discussion.
IRIS is a NASA small explorer mission developed and operated by LMSAL with mission operations executed at NASA Ames Research center and major contributions to downlink communications funded by ESA and the Norwegian Space Centre.
Courtesy of NASA/SDO, the AIA and HMI teams and JSOC.

\section{Appendix: Network jet identification and tracking algorithm}
The network jets are identified and traced on the IRIS SJ 1330\,\AA\ images by the automatic method.
The algorithm of the method includes a few steps as described in the follow.

\par
(a) Region selection and base definition:
In order to simplify the tracking procedures, we analyze the regions ``R1''--``R4'', where the bright elements in the network lanes (viewed in IRIS SJ 1330\,\AA) are almost aligned and the network jets are ejected in directions almost parallel to each other.
The bases of the network jets in each region are determined to be the edge of the bright network lanes.
Because the network lanes are very dynamic with many transient bright dots, we make an artificial image that is sum of all the IRIS SJ 1330\AA\ images taken in the observing period of time, and the edge of the bright network lanes are defined based on the artificial image.

\par
\begin{figure*}[!ht]
\centering
\includegraphics[clip,trim=0cm 0cm 0cm 0cm,width=\textwidth]{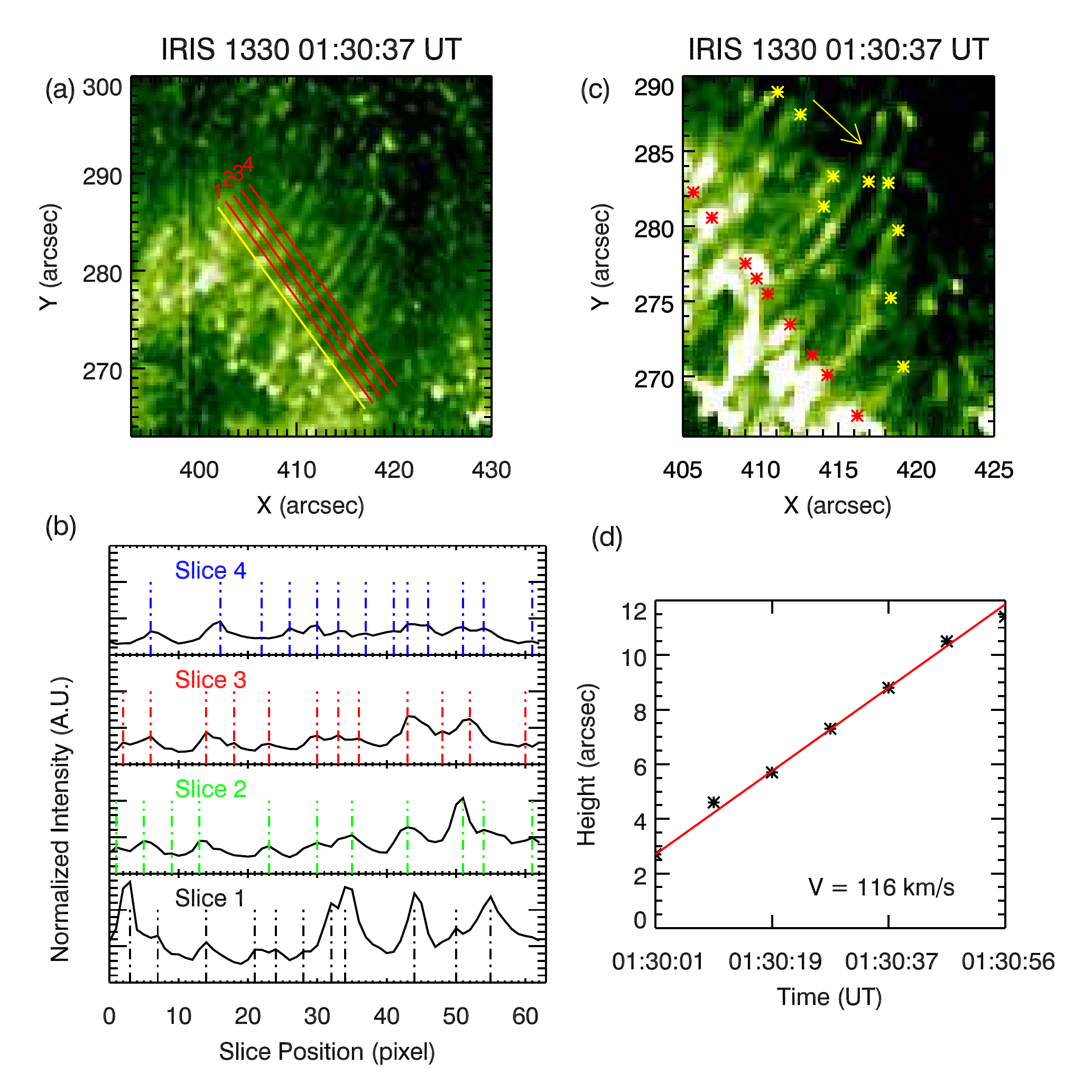}
\caption{The procedures of the automatic network jet identification and tracking algorithm.
(a) A network region (``R1'') observed in IRIS SJ 1330\,\AA.
The yellow line indicates the base of the network jets originated in this network.
The slices (red lines denoted by ``1--4'') are used to identified jet-like features (see the text for details).
(b) The IRIS SJ 1330\,\AA\ radiation variations along the slices 1--4. The local peaks identified in these variations are denoted by dash-dotted lines.
(c) Examples of identified jet-like features on an image frame at 01:30:37\,UT. The red asterisks mark the bases of the features and the yellow asterisks mark the top of them.
The arrow point to a network jet that shown as example in panel (d).
(d) The heights of the network jet (denoted in panel c) varying with time.
The speed of the network jet is obtained by the slope of the linear fit of this variation (red line).}
\label{figmethod1}
\end{figure*}

(b) Jet-like feature identification in an image frame:
In each region, starting from the bases of the network jets, we select four slices that extend (almost) perpendicular to the network jets and have 1\arcsec\ separation between each two neighbours (see Figure\,\ref{figmethod1}a).
The variation of the SJ 1330\,\AA\ radiance along each slice is then obtain.
The local peaks as defined in \citet{2017MNRAS.464.1753H} are identified on the variation curves (see Figure\,\ref{figmethod1}b).
Such a local peak indicates a local brightening.
In order to avoid the noise effect, the local peaks with radiances less than 2.5 times of the background are excluded.
If a local peak of a slice and the other local peak of the neighbouring slice appear at the slice positions with difference less than 1\arcsec,
they are considered to be results of the same bright feature.
If a bright feature as defined by local peaks is found in all the four slices,
it is considered to be a jet-like feature, and its location and extending direction are determined using the positions of the corresponding local peaks identified in the four slices.

\par
(c) Length determination for the jet-like features:
For each identified jet-like feature, a slice along its extending direction is made and the radiance variation along the slice is obtained.
Along the slice, the farthest extending point of the jet-like feature is defined to be the location where the radiance drops below 2.5 times of the background (See Figure\,\ref{figmethod1}c).

\par
\begin{figure*}[!ht]
\centering
\includegraphics[trim=0cm 0cm 0cm 20cm,clip,width=\textwidth]{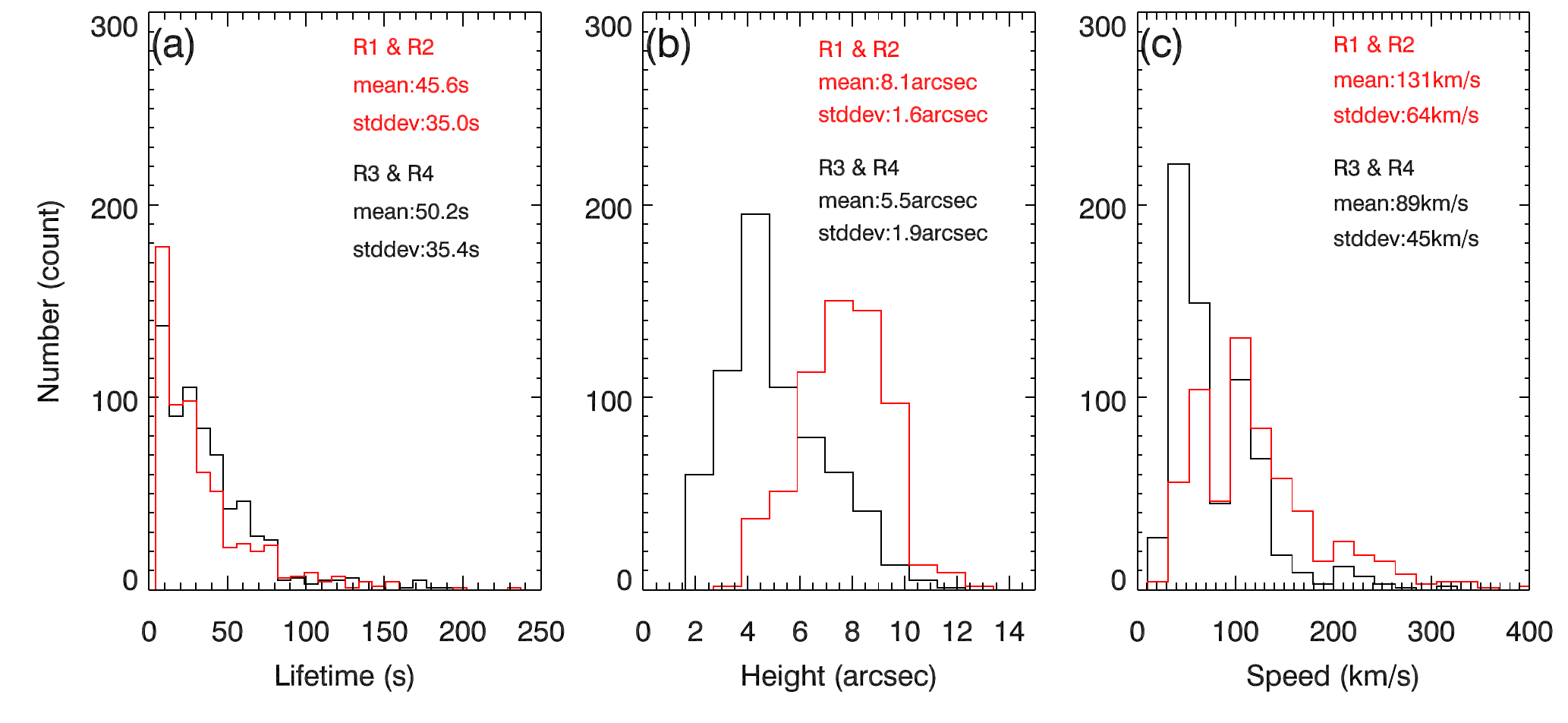}
\caption{The statistical histograms of the lifetimes (a), heights (b) and speeds (c) of the network jets identified and tracked in the regions of ``R1\&R2'' (red lines) and ``R3\&R4'' (black lines).}
\label{figstatres}
\end{figure*}

(d) Jet-like feature tracking and definition of network jet and its lifetime and speed:
Steps (b) and (c) is run for each imaging frame.
If a jet-like features found in one frame and the other jet-like feature found in the other close-in-time frame locate at the same position,
they are considered to be the same jet-like feature that is evolving in time.
If a jet-like feature is found in more than one image frame and less than 27 image frames (i.e. lifetime less than 4 minutes), it is considered to be a network jet.
The heights of the network jet seen at different times are given by the lengths of the jet-like features.
By following the heights of a network jet at different times, its speed could be obtained using a linear fit in the height vs. time space (see Figure\,\ref{figmethod1}d).
Please note that for each jet only its maximum height is used in the statistics (i.e. Figure\,\ref{figstatres}b).

\bibliographystyle{spr-mp-sola}
\bibliography{youqian}

\end{article}

\end{document}